\documentclass{emulateapj}
\usepackage{amssymb}
\usepackage{lscape}

\begin{document}
\newcommand{\lya}{Lyman~$\alpha$}
\newcommand{\lyb}{Lyman~$\beta$}
\newcommand{\degpoint}{\mbox{$^\circ\mskip-7.0mu.\,$}}
\newcommand{\minpoint}{\mbox{$'\mskip-4.7mu.\mskip0.8mu$}}
\newcommand{\secpoint}{\mbox{$''\mskip-7.6mu.\,$}}
\newcommand{\sqdeg}{\mbox{${\rm deg}^2$}}
\newcommand{\squig}{\sim\!\!}
\newcommand{\subsun}{\mbox{$_{\twelvesy\odot}$}}
\newcommand{\et}{{\it et al.}~}
\newcommand{\Rs}{{\cal R}}

\def\ltsima{$\; \buildrel < \over \sim \;$}
\def\simlt{\lower.5ex\hbox{\ltsima}}
\def\gtsima{$\; \buildrel > \over \sim \;$}
\def\simgt{\lower.5ex\hbox{\gtsima}}
\def\propsima{$\; \buildrel \propto \over \sim \;$}
\def\simprop{\lower.5ex\hbox{\propsima}}
\def\arcs{$''~$}
\def\arcm{$'~$}

\title{A POSSIBLE CORRELATION BETWEEN THE LUMINOSITIES AND LIFETIMES OF ACTIVE GALACTIC NUCLEI\altaffilmark{1}}

\author{\sc Kurt L. Adelberger\altaffilmark{2}}
\affil{Carnegie Observatories, 813 Santa Barbara St., Pasadena, CA, 91101}

\author{\sc Charles C. Steidel}
\affil{Palomar Observatory, Caltech 105--24, Pasadena, CA 91125}

\altaffiltext{1}{Based, in part, on data obtained at the W.M. Keck
Observatory, which is operated as a scientific partnership between
the California Institute of Technology, the University of California,
and NASA, and was made possible by the generous financial support
of the W.M. Keck Foundation.}
\altaffiltext{2}{Carnegie Fellow}

\begin{abstract}
We use the clustering of galaxies around distant AGN to show 
with $\sim 90$\% confidence that
fainter AGN are longer lived.  Our argument is simple:
since the measured galaxy-AGN cross-correlation length $r_0\sim 5h^{-1}$ Mpc does not
vary significantly over a 10 magnitude range in AGN optical luminosity,
faint and bright AGN must reside in dark matter halos with similar masses.
The halos that host bright and faint AGN therefore have
similar intrinsic abundances, and the large observed variation in AGN number
density with luminosity reflects a change in duty cycle.
\end{abstract}

\submitted{Received 2005 March 10; Accepted 2005 May 11 }
\shorttitle{LUMINOSITIES AND LIFETIMES OF AGN}
\shortauthors{Adelberger \& Steidel}
                                                                                
\section{INTRODUCTION}
\label{sec:intro}
In the famous paper that
postulated a link
between quasars and accreting black holes, 
Lynden-Bell (1969) remarked that the black holes
created by quasar accretion would be gigantic
and common, with masses around $10^8M_\odot$ and
a space density similar to that of local galaxies.
It was a prescient comment, but
So\/ltan's (1982) refinement of his calculation 
drew attention to the importance of the assumed
quasar lifetime.  The total accretion was sufficient
to place a $10^6M_\odot$ black hole inside every
galaxy brighter than M31, So\/ltan showed,
but the accreted mass might 
equally well be distributed among a smaller number
of heavier black holes or a larger number of lighter ones.
The length $t_Q$ of the quasars' lives would determine
which was the case.  
Although the understanding of black hole formation has
advanced enormously since that time,
$t_Q$ remains a key parameter in theoretical models.
Our ignorance of it is arguably
the largest source of uncertainty in the accretion histories
of supermassive black holes.

This paper is concerned not with the value of the quasar lifetime itself,
but rather with the idea that there is a single
lifetime for accretion onto active galactic nuclei (AGN).
It is obviously an oversimplification.
The duration of a luminous accretion episode is presumably
affected by the mass of the central black hole, the size of the gas
supply, the nature of the event that funnels gas towards the black hole,
the strength and duration of dust obscuration,
and so on.  
Our aim is to measure the extent to which this produces
a systematic dependence of the lifetime on the luminosity of the AGN.

It is easy to convince oneself that such a dependence might exist.
The extreme accretion associated
with the most luminous QSOs is rare and must have
a small duty cycle (e.g., Martini 2004),
while low-level accretion has a high enough duty cycle
to be observed in approximately half of all
nearby galaxies (e.g., Ho 2004).
As far as we know, 
however, no-one has previously attempted a
direct measurement of
the dependence of AGN lifetime on luminosity
(cf. Merloni 2004, Hopkins et al. 2005).
Although it may seem perverse to try to look for
systematic differences in the accretion lifetime
when the lifetime is still uncertain by two orders
of magnitude (e.g., Martini 2004),
in fact (as we show in \S~\ref{sec:methods}) changes in the
lifetime are much easier to measure than the
value of the lifetime itself.

Our approach exploits the well known fact
that the duty cycle
of a population of objects can be inferred from its number density
and clustering strength
(e.g., Adelberger et al. 1998).  The reason is simple.  In universes
with hierarchical structure formation, the rarest and most massive
virialized halos cluster the most strongly (e.g., Kaiser 1984),
and so the mass and number density of the sub-population of halos that contain the objects
can be deduced from the strength of the objects' clustering.
The duty cycle is equal to the objects' observed number density
divided by the number density of halos that can host them.  
If clustering measurements
indicate that AGN reside in halos of mass $10^{12}M_\odot$, for example,
but the number density of AGN is only 1\% of the number density
of halos with $M=10^{12}M_\odot$, the duty cycle is evidently 0.01.

Martini \& Weinberg (2001) and Haiman \& Hui (2001) were the first
to discuss the technique in detail.
Our treatment is similar to theirs, except
in one important respect:  we
infer the duty cycle from the clustering of galaxies around AGN, rather than
from the clustering of AGN themselves.  As pointed out by
Kauffmann \& Haehnelt (2002), the high number density of galaxies
makes the galaxy-AGN cross-correlation length much easier to
measure than the AGN auto-correlation length.  A major additional
benefit is that any survey deep enough to detect galaxies
around bright high-redshift QSOs will inevitably detect
faint AGN at the same redshifts, increasing the sample size
and the luminosity baseline over which changes in duty cycle
can be measured.

\section{DATA}
\label{sec:data}
\subsection{Galaxies}
The data we analyzed were taken from our color-selected surveys of
star-forming galaxies with magnitude ${\cal R}_{\rm AB}\leq 25.5$
and redshift $1.8\simlt z\simlt 3.5$.
A more complete description of the surveys can be found in Steidel et al. (2003),
Steidel et al. (2004), and Adelberger et al. (2005b).  We review only the
most important aspects here.

\begin{figure}
\plotone{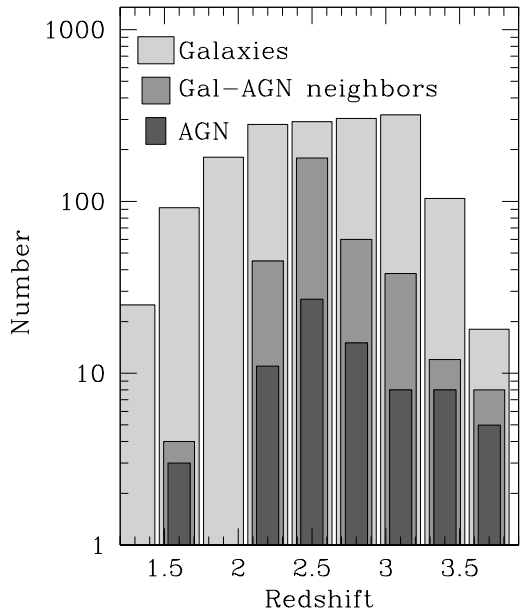}
\caption{
Redshift distributions for the galaxies and AGN in our sample.
Also shown is the number of galaxy-AGN neighbors, defined as
the number of galaxy-AGN pairs with angular separation
$60''<\theta<300''$ [$1.2\simlt R/(h^{-1}{\rm comoving\, Mpc})\simlt 6.2$]
and radial separation $\Delta Z<30h^{-1}$ comoving Mpc.
\label{fig:nz}
}
\end{figure}

\begin{figure}
\plotone{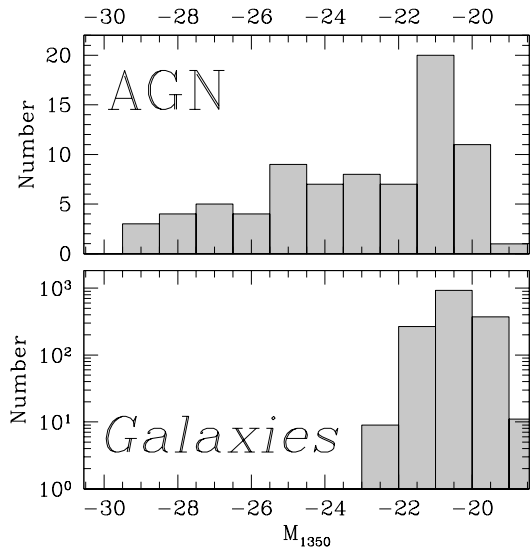}
\caption{
Distribution of absolute AB magnitude at rest-frame $1350$\AA\
for the AGN and galaxies in our spectroscopic sample.  
No corrections for
incompleteness have been applied, so these do not resemble the
true distributions for the underlying populations.
\label{fig:M1350hist}
}
\end{figure}

The surveys consist of measured redshifts for 1627 galaxies 
with redshift $z>1$ in $19$ fields
scattered around the sky (table~\ref{tab:fields}).  
(These totals exclude any survey fields
with no detected AGN and include only the galaxies with
the most certain redshifts.)  The size of the fields varies but is typically
$\sim 100$--$200$ square arcmin.  The coordinates of some fields were chosen
more-or-less at random, but most fields were centered on a bright QSO
or group of QSOs.  Objects were selected for spectroscopy if their
$U_nG{\cal R}$ colors indicated they were likely to lie in the targeted range
of redshifts.  Our decision to obtain a spectrum of an object was influenced
only by its $U_nG{\cal R}$ colors, ${\cal R}$ magnitude, and spatial position;
we were more likely to observe objects if they had $23<{\cal R}<24.5$,
if they had colors similar to those expected for AGN, or if they lay close
to a known AGN, and we rarely observed objects
whose colors did not satisfy the selection
criteria of Steidel et al. (2003) and Adelberger et al. (2004).
The overall redshift distribution of the galaxies in these fields is shown in
figure~\ref{fig:nz}.  Their distribution of absolute magnitudes, calculated
from observed broadband colors
for a concordance cosmology with $\Omega_M=0.3$, $\Omega_\Lambda=0.7$, $h=0.7$,
is shown in figure~\ref{fig:M1350hist}.

\begin{deluxetable*}{lrrrrr}
\tablewidth{0pc}
\scriptsize
\tablecaption{Observed fields}
\tablehead{
        \colhead{Field} &
        \colhead{$\alpha(2000)$} &
        \colhead{$\delta(2000)$} &
        \colhead{$N_{\rm gal}$\tablenotemark{a}} &
        \colhead{$N_{\rm AGN}^{>-24}$\tablenotemark{b}} &
        \colhead{$N_{\rm AGN}^{<-24}$\tablenotemark{c}} 
}
\startdata
B20902+34        & 09 05 31 &  34 08 02           &    31 &     1 &     0\\
CDFb             & 00 53 42 &  12 25 11           &    19 &     1 &     0\\
DSF2237a         & 22 40 08 &  11 52 41           &    41 &     1 &     0\\
DSF2237b         & 22 39 34 &  11 51 39           &    43 &     2 &     1\\
HDF              & 12 36 51 &  62 13 14           &   251 &     5 &     1\\
Q0000-263\tablenotemark{d} & 00 03 23 & -26 03 17           &    15 &     2 &     0\\
PKS0201+113      & 02 03 47 &  11 34 45           &    23 &     1 &     1\\
LBQS0256-0000    & 02 59 06 &  00 11 22           &    45 &     2 &     1\\
LBQS0302-0019    & 03 04 50 &  00 08 13           &    42 &     1 &     1\\
FBQS J0933+2845  & 09 33 37 &  28 45 32           &    63 &     1 &     1\\
Q1305            & 13 07 45 &  29 12 51           &    76 &     4 &     3\\
Q1422+2309       & 14 24 38 &  22 56 01           &   108 &     5 &     1\\
Q1623            & 16 25 45 &  26 47 23           &   200 &     9 &     7\\
HS1700+6416      & 17 01 01 &  64 12 09           &    88 &     1 &     1\\
Q2233+136        & 22 36 27 &  13 57 13           &    43 &     3 &     1\\
Q2343+125        & 23 46 05 &  12 49 12           &   188 &     2 &     4\\
Q2346            & 23 48 23 &  00 27 15           &    44 &     3 &     3\\
SSA22a           & 22 17 34 &  00 15 04           &    59 &     0 &     2\\
WESTPHAL         & 14 17 43 &  52 28 48           &   248 &     7 &     0\\
Total:           &          &                     &  1627 &    51 &    28\\
\enddata
\tablenotetext{a}{Number of (non-active) galaxies with spectroscopic redshift $z>1$.}
\tablenotetext{b}{Number of AGN with spectroscopic redshift $z>1$ and rest-frame $1350$\AA\ absolute AB
magnitude $M_{1350}>-24$}
\tablenotetext{c}{Number of AGN with spectroscopic redshift $z>1$ and $M_{1350}\leq -24$}
\tablenotetext{d}{The field is centered on this QSO, but the QSO itself is excluded from our analysis because
we lack a good spectrum.}
\label{tab:fields}
\end{deluxetable*}


\subsection{AGN}
Fifty-seven of the 1684 objects in our spectroscopic sample have strong emission
in both Lyman-$\alpha$ and CIV $\lambda 1549$.  
We classify these objects as AGN for reasons that are discussed in Steidel et al. (2002).
Although some of our faintest AGN might be misclassified as galaxies because
their CIV lines are too weak for us to detect, the lack of CIV emission in
the thousand-object composite spectrum of Shapley et al. (2003) shows that
these misclassified AGN must be rare.

\begin{figure}
\plotone{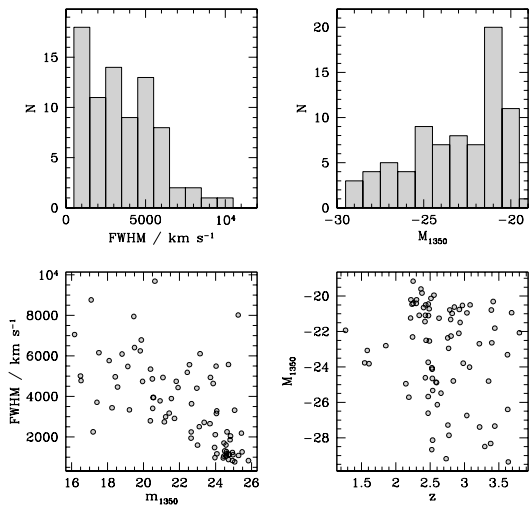}
\caption{
Overview of the characteristics of the AGN in our sample.
{\it Upper left:} Histogram of CIV line width.  The typical
uncertainty ranges from 10--20\%, and is dominated by
systematics (e.g., continuum placement) for the brightest AGN.
{\it Upper right:} Histogram of absoluate AB magnitude
at rest-frame 1350\AA\ ($M_{1350}$).  The uncertainty in the AB magnitude
is $\simlt 0.2$ magnitudes for even our faintest objects (e.g., 
Steidel et al. 2003).
{\it Lower left:} Relationship between CIV line width
and apparent AB magnitude at rest-frame $1350$\AA.
{\it Lower right:} $M_{1350}$ against redshift.
Recall that the selection bias is severe in our AGN sample, 
since (for example) we
deliberately targeted AGN that were bright and had broad emission lines.
These panels show the characteristics of our sample as selected, not
of a fair sample of high-redshift AGN.
\label{fig:check_mbh}
}
\end{figure}

Our total sample of AGN was increased to 79 by adding the previously known AGN
that we deliberately included in our survey fields.  
Since CIV was the only line (aside from Lyman-$\alpha$) detected with reasonable
significance in every AGN spectrum, we based our redshift assignments on it.
In their analysis of 3814 QSOs from the Sloan Digital Sky Survey,
Richards et al. (2002) found that CIV was blueshifted on average by 824 km s$^{-1}$
compared to MgII, which they assumed was at the QSO's systemic redshift.
We accordingly assumed that each of our AGN's true redshift was
824 km s$^{-1}$ redder than the peak of CIV emission.  Since Richards et al. (2002)
report a scatter in the CIV--MgII velocity offsets of $500$ km s$^{-1}$, we expect
that the uncertainty in our QSO redshifts will be approximately $500$ km s$^{-1}$.
Although the way we assign redshifts is better suited to our sample's 
broad-lined AGN, any mistakes in the redshifts of narrow-lined AGN
are unlikely to affect our conclusions:
as we will see, the typical redshift error would have to be
$\sim 3000$ km s$^{-1}$ (i.e., $\sim 30 h^{-1}$ comoving Mpc)
to alter our clustering measurements significantly.
Figure~\ref{fig:nz} shows the 
redshift distribution for the 79 AGN.  
Figure~\ref{fig:check_mbh} shows their distribution of
velocity FWHM and apparent magnitude.

Strong emission lines
prevented us from calculating the AGNs'
AB magnitudes at rest-frame 1350\AA\ directly from their broadband magnitudes.
Instead we scaled
each AGN's spectrum to match its observed $G$
and ${\cal R}$ magnitudes, measured the flux density near 1350\AA,
then converted to
absolute magnitude for a cosmology with $\Omega_M=0.3$, $\Omega_\Lambda=0.7$, $h=0.7$.
This procedure failed for our brightest sources, $G\simlt 18$,
which were saturated in our images.  For these we adopted
the magnitude implied by their unscaled flux-calibrated spectra.
Three of our sources were saturated and lacked flux-calibrated spectra.  The magnitudes
of these were taken from the Sloan Digital Sky Survey archive or from 
photographic measurements in the
NASA/IPAC Extragalactic Database.
Figure~\ref{fig:M1350hist} shows the resulting histogram of AGN absolute magnitude.
Although unintended, our selection strategy has given us
a sample of AGN with brightnesses distributed almost uniformly over a 10-magnitude range.
Comparison to the galaxies' apparent magnitude distribution suggests
that stellar light may contribute significantly to the measured magnitudes
of the faintest AGN.  We do not correct for this.  Doing so would only strengthen 
our conclusions, since the faintest AGN would be even fainter than we
assume.

\subsection{Simulations}
In a number of places our interpretation of the data
relies on the GIF-LCDM numerical simulation of structure formation in
a cosmology 
with $\Omega_M=0.3$, $\Omega_\Lambda=0.7$,
$h=0.7$, $\Gamma=0.21$, and $\sigma_8=0.9$.
This gravity-only simulation
contained $256^3$ particles with mass $1.4\times 10^{10} h^{-1} M_\odot$
in a periodic cube of comoving side-length $141.3h^{-1}$ Mpc,
used a softening length of $20 h^{-1}$ comoving kpc, and
was released publicly, along with its halo catalogs, by
Frenk et al. astro-ph/0007362.  Further details
can be found in Jenkins et al. (1998) and Kauffmann et al. (1999).
Although the simulation does not include much of the physics associated
with galaxy formation, we make use only of its predictions for the 
statistical distribution of dark matter on large ($\simgt$ Mpc) scales.  
Since the GIF-LCDM cosmology is consistent
with the Wilkinson Microwave-Anisotropy Probe results
(Spergel et al. 2003), and since modeling the gravitational
growth of perturbations
on large scales is not numerically challenging,
the large-scale distribution of dark matter in this simulation should
closely mirror that in the actual universe.

\section{METHODS}
\label{sec:methods}
\subsection{Estimating $r_0$}
\label{sec:r0methods}

We estimated the correlation lengths of the samples with two
approaches.  Both correct for the irregular angular sampling of our spectroscopy
and are unaffected by the selection criteria that were used to include AGN 
in our sample.  The second approach is also insensitive to the criteria
that were used to select the galaxies.  See Adelberger (2005) for a more
complete discussion.

In the first approach, we cycle through the AGN in our sample, calculating
for each one both the number $N_{\rm obs}(\ell)$ of galaxies
in the AGN's field
whose comoving radial separation from the AGN, $\Delta Z$, is less than
$\ell=30h^{-1}$ Mpc, and the number $N_{\rm exp}(\ell,r_0)$ that
would be expected if the correlation function had the form
$\xi(r) = (r/r_0)^{-1.6}$.  The quantity $N_{\rm exp}(\ell,r_0)$ is
related straightforwardly to the integral of the correlation function along the lines
of sight to galaxies in the field.  As shown by Adelberger (2005),
\begin{equation}
N_{\rm exp}(\ell,r_0) = \sum_j^{\rm galaxies}\frac{\int_{z_i-\Delta z}^{z_i+\Delta z} dz\, P_j(z) [1+\xi(r_{ij})]}{\int_{0}^{\infty} dz\, P_j(z) [1+\xi(r_{ij})]}
\label{eq:barn}
\end{equation}
where the sum runs over all galaxies in the AGN's field,
$z_i$ is the AGN's redshift,
$\Delta z$ is the redshift difference corresponding to a comoving radial separation of 
size $\ell$,
$P_j(z)$ is the selection function for the $j$th galaxy\footnote{
Since the galaxies in our samples were chosen with different color-selection
criteria, their expected redshift distributions are different.
In this approach, we set $P_j$ to the observed LBG redshift distribution
if the object was selected with the LBG selection criteria and to
the observed BX redshift distribution if the object was selected with the BX
criteria.  Otherwise the galaxy is ignored.
(See Adelberger et al. 2004 for a definition of these criteria
and plots of their redshift distributions.)  
}, normalized
so that $\int_0^\infty dz\, P_j(z) = 1$,
and $r_{ij}$ is the distance between the AGN 
and a point at redshift $z$ with the galaxy's angular separation $\theta_{j}$.
We then sum the values $N_{\rm obs}(\ell)$ and $N_{\rm exp}(\ell,r_0)$ for all our AGN,
and take as our best-fit correlation length the value of $r_0$
that makes the total expected neighbor counts equal to the total observed.
To ensure that our estimate of $r_0$ reflects the clustering strength
on large ($\simgt$ Mpc) scales, rather than conditions inside the AGNs' halos,
we exclude from consideration any galaxy-AGN pairs with angular separation
$\theta<60''$ (i.e., $1.2h^{-1}$ comoving Mpc at $z=2.5$).  Galaxy-AGN pairs
with $\theta>300''$ are also excluded,
since the weak clustering signal at the largest angular separations
can be overwhelmed by low-level systematic errors (Adelberger 2005).

The approach of the preceding paragraph
can fail if the assumed selection functions $P_j$ are inaccurate.
To guard against this possibility, we also estimate $r_0$ by finding
the value that makes 
\begin{equation}
\frac{\sum_{\rm AGN}N_{\rm obs}(\ell)}{\sum_{\rm AGN}N_{\rm obs}(2\ell)} = \frac{\sum_{\rm AGN}N_{\rm exp}(\ell,r_0)}{\sum_{\rm AGN}N_{\rm exp}(2\ell,r_0)}
\label{eq:defk}
\end{equation}
Taking the ratio causes most systematic errors to cancel (Adelberger 2005).
Since it also increases the random errors, however, we use
the equation~\ref{eq:defk} only to verify
that systematic errors have not badly compromised the estimate of $r_0$
from the first approach.

\subsection{Estimating the duty cycle}
\label{sec:estimatedutycycle}
As stated in the introduction, our definition of duty cycle
is the observed number density of AGN divided by the number density
of halos that can host them.  Calculating it requires two steps.

\subsubsection{Halo abundance}
\begin{figure}
\plotone{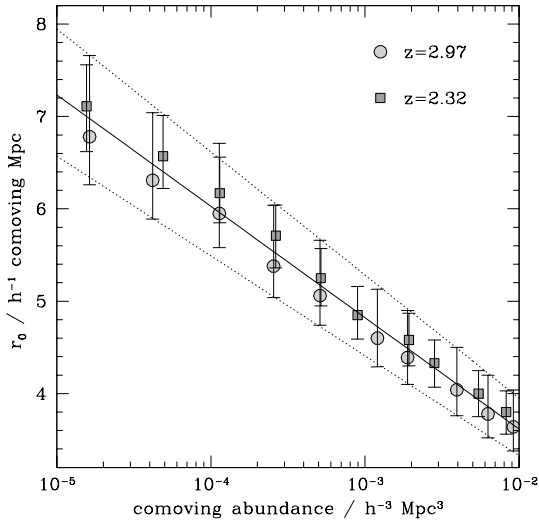}
\caption{
Theoretical relationship between cross-correlation length $r_0$ and AGN-halo
comoving abundance $n$.  Points show the GIF-LCDM relationship at two redshifts.
The error bars indicate the uncertainty in the relationship due to
the uncertainty in the galaxies' threshold mass.
The solid line shows the least-squares compromise that we adopt throughout:
$\log(n/(h^{-1}{\rm Mpc})^3) = -0.83 r_0 + 1.00$.
The upper and lower dotted lines show the relationships that would
result if we altered the assumed threshold mass by $\pm 1\sigma$.
Figure~\ref{fig:r0dutycycle} shows that our conclusions would not
be significantly affected if we adopted these relationships instead.
\label{fig:r0_vs_abundance}
}
\end{figure}

We use the GIF-LCDM simulations to estimate the halo abundance 
from $r_0$.  For each of the publicly released catalogs of halos
at redshifts $2<z<3$\footnote{i.e., for the
catalogs at $z=2.97$, 2.74, 2.52, 2.32, 2.12}, we calculated the cross-correlation function
$\xi_{M_1,M_2}(r)$ of
halos in two mass ranges, $M>M_1$ and $M>M_2$, for different choices
of $M_1$ and $M_2$, and estimated the cross-correlation length
$r_0$ by fitting a power-law to $\xi_{M_1,M_2}$
at separations $1<r/(h^{-1}{\rm Mpc})<10$.
After calculating the number density of halos with $M>M_2$ in
the simulations at redshift $z$, we stored our results
as a table $r_0(z,M_1,n_2)$ giving the expected cross-correlation
length at redshift $z$ between halos with mass threshold $M_1$ and halos with
number density $n_2$.  If all our observations were at
redshift $z_0$ and we knew the threshold mass $M_g$ of the galaxies' halos,
we could convert any measured correlation length $r_0$ into
a number density $n_q$ of AGN halos by simply looking up the value
of $n_q$ that made the tabulated $r_0(z_0,M_g,n_q)$ equal our
observed correlation length.  In fact our observations are at a range
of redshifts and the galaxy mass is not precisely known.  
Figure~\ref{fig:r0_vs_abundance} shows the uncertainty in
the relationship between $r_0$ and $n_q$ that results from
the range of redshifts in our survey and from the $1\sigma$
uncertainty in the galaxy masses (Adelberger et al. 2005a).
For the remainder of the paper
we will adopt an $r_0$--$n_q$ relationship that is a least-squares
fit to the data in the figure (solid line).
Although we can offer little justification
for this compromise, the exact choice of relationship has almost
no effect on our conclusions.  Any errors in the relationship
increase or decrease in tandem the implied duty cycles
for bright and faint AGN; they alter the absolute value
we infer for the duty cycles but not the relative difference
between them.  (We demonstrate that this is true
in Figure~\ref{fig:r0dutycycle}, below.)  This is one of the main strengths
of our approach.  It justifies our claim in \S~\ref{sec:intro}
that a systematic variation of AGN lifetime with luminosity is
easier to measure than the
absolute value of the lifetime itself.

\subsubsection{AGN abundance}
\label{sec:agndensity}
\begin{figure}
\plotone{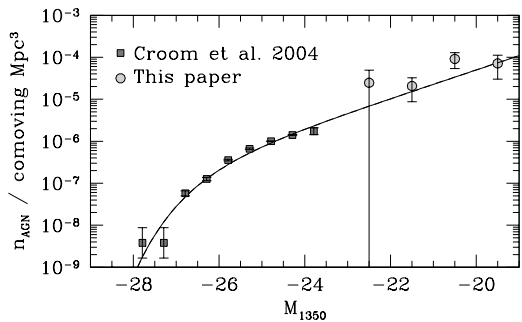}
\caption{
Observed number density vs.~magnitude for AGN at $z\sim 2$.  Squares
show the 2dF QSO luminosity function of Croom et al. (2004).  Circles
show our rough estimate of the AGN luminosity function at fainter 
magnitudes, calculated from our survey with the method described
in \S~\ref{sec:agndensity}.  The crude completeness corrections of
this approach yield a luminosity function adequate only
for cases like ours where low accuracy is tolerable.
The parameters of the Schechter-function
(solid line; $M_\ast = -26.2$, $\alpha=-1.85$, 
$\Phi_\ast=4\times 10^{-7}$ Mpc$^{-3}$)
should not be used in other situations.
\label{fig:agndensity}
}
\end{figure}

We adopt a crude approach since small (tens of percent) errors in
the AGN abundance have little effect on our conclusions.  
At the faintest magnitudes we estimate the AGN number density by
multiplying the galaxy luminosity function at $z=3$ (Adelberger \& Steidel 2000)
by $f(M_{1350})$, the fraction of sources in our spectroscopic sample
with absolute magnitude $M_{1350}$ that were observed to be AGN.
Note that we are including all AGN in this analysis, not merely
the broad-lined AGN considered by Hunt et al. (2004).
Since the faint end of the rest-frame UV luminosity distribution
of galaxies does not evolve significantly from $z=3$ to $z=2$
(N.A. Reddy et al. 2005, in preparation), this number density
should be roughly appropriate down to $z=2$.
At the brightest magnitudes we adopt the ``2dF'' $1.81<z<2.10$
QSO luminosity function of Croom et al. (2004).\footnote{We
convert the absolute magnitudes $M_{b_J}$ reported
by Croom et al. (2004) to $M_{1350}$ by adding 0.46 magnitudes;
subtracting 0.07 magnitudes converts to the AB system,
and adding 0.53 undoes their $K$-correction from observed-frame to rest-frame $b_J$
(Cristiani \& Vio 1990).}
The AGN luminosity distribution is fit tolerably well by a Schechter
function (figure~\ref{fig:agndensity}), and we use this fit to
estimate the number density of AGN in each range of apparent magnitude.

\section{RESULTS}
\label{sec:results}
\begin{figure}
\plotone{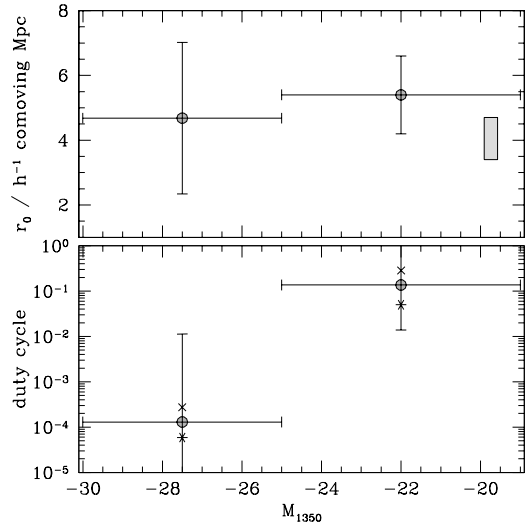}
\caption{
Top panel:  galaxy-AGN cross-correlation length as a function of
AGN luminosity $M_{1350}$.  Points with error bars show our measurements.
The shaded square shows the $\pm 1\sigma$ range of the galaxy-galaxy
correlation length at similar redshifts (Adelberger et al. 2005a);
its abscissa is arbitrary.  Bottom panel:  implied duty cycle as a 
function of AGN luminosity.  
Error bars show the $\pm 1\sigma$ random uncertainty.
The four- and six-pointed stars show how our estimated duty cycle
would change if we altered the assumed relationship between
clustering strength and abundance by an amount similar to its
uncertainty.
(They correspond to the upper and lower dotted envelopes in 
Figure~\ref{fig:r0_vs_abundance}.)
Note that the confidence intervals shown in this plot
reflect only the constraints from our clustering analysis. Other
considerations rule out a duty cycle of $\simgt 1$
for the faint AGN and $\simlt 10^{-5}$ for the bright AGN, however.
See \S~\ref{sec:summary} for further discussion.
\label{fig:r0dutycycle}
}
\end{figure}

The first approach of \S~\ref{sec:r0methods} leads to the
estimates $r_0=4.7,5.4$ comoving Mpc for the galaxy-AGN cross-correlation length
of AGN with magnitude $-30<M_{1350}<-25$ and $-25<M_{1350}<-19$,
respectively.
An easy way to estimate the uncertainty is suggested by
the similarity of the cross-correlation length to the galaxy-galaxy
correlation length reported by Adelberger et al. (2005a):
generate many alternate realizations of the data by 
treating randomly chosen galaxies in each field as that
field's AGN, rather than the true AGN themselves,
and recalculate $r_0$ 
for each simulated sample.  The rms dispersion of $r_0$ among
these simulated samples should be roughly equal to the
uncertainty in $r_0$, and we adopted it for the error
bars in the top panel of Figure~\ref{fig:r0dutycycle}.
The true uncertainty is likely to be somewhat smaller,
since our spectroscopic selection strategy gave our AGN
more angular neighbors with measured redshifts than the typical galaxy.

The bottom panel of Figure~\ref{fig:r0dutycycle} shows
the same data, except 
the cross-correlation length has been converted to
a duty cycle with the approach of \S~\ref{sec:estimatedutycycle}.
As emphasized in that section, uncertainties in the $r_0$--abundance
relationship mean that the labels on the $y$ axis could be wrong
by a multiplicative constant, but relative differences in
duty cycle should be secure.  

To estimate the significance of the apparent difference in duty cycle,
we note that our adopted relationship between
$r_0$ and halo number density implies that
$r_0$ would be $2.92h^{-1}$ comoving Mpc larger
for AGN with $-30<M_{1350}<-25$ than AGN with $-25<M_{1350}<-19$
under the null hypothesis that the duty cycle is
independent of $M_{1350}$.  The observed difference
in best-fit correlation length, $-0.72h^{-1}$ comoving Mpc,
is therefore $3.64h^{-1}$ comoving Mpc smaller than
the difference that would be expected under the null hypothesis.
A difference as large or larger than
$\Delta r_0=3.64h^{-1}$ Mpc between AGN with
$-25<M_{1350}<-19$ and $-30<M_{1350}<-25$
occurred in 10\% of the randomized
AGN samples described above.  We conclude that the
null hypothesis of a constant duty cycle can be rejected with
roughly 90\% confidence.

\section{SUMMARY \& DISCUSSION}
\label{sec:summary}
We measured the galaxy-AGN cross-correlation length $r_0$
as a function of AGN luminosity.  The cross-correlation length
was similar for bright and faint AGN,
$r_0=4.7\pm 2.3$ for $-30<M_{1350}<-25$ and $r_0=5.4\pm 1.2$ for $-25<M_{1350}<-19$,
which led us to conclude with $90$\% confidence
that both are found in halos with similar masses
and that bright AGN are rarer because their duty cycle is shorter.
Since halo lifetimes depend only weakly on halo
mass (e.g., Martini \& Weinberg 2001), the difference in duty cycle
implies that optically faint AGN have longer lifetimes.

Our analysis differs from previous work (e.g., that of Croom et al. 2005,
who also found no luminosity dependence in the AGN clustering strength)
in two principal ways.  We estimated the duty cycle from the cross-correlation
of galaxies and AGN, not from the auto-correlation function of
AGN, and our sample included AGN with a much wider range
of luminosities, extending $\sim 4$ magnitudes fainter than
the QSO threshold $M_{1350}=-23$.  These differences allowed
us to obtain our measurement from a comparatively small survey.

An appraisal of this result should cover at least the
following three points.

The first is obvious:  it
is only marginally significant.  
Larger samples will be required to prove that
the duty cycle depends on luminosity.  
Moreover, other arguments suggest that the minimum allowed duty cycle
at high luminosity should be increased and the maximum allowed at low luminosity
should be decreased.  Since the AGN lifetime is roughly
the age of the universe times the duty cycle (e.g., Martini \& Weinberg 2001),
a duty cycle of $\simlt 10^{-5}$ for the brightest AGN is incompatible
with the observed proximity effect in QSOs' spectra (e.g., Martini 2004)
and with
the lack of flickering QSOs in the Sloan Digital Sky Survey 
(Martini \& Schneider 2003).
A duty cycle of roughly unity for the fainter AGN is implausible as well,
since a black hole radiating continuously would almost certainly
be too faint compared to its galaxy for us to detect:
the difference in energetic efficiency for black hole
accretion ($0.1mc^2$) and hydrogen burning ($0.007mc^2$)
implies that 
a galaxy's steadily radiating black hole would be much fainter than its
stars if the final ratio of black hole to stellar mass is
$M_{\rm BH}/M_\ast\sim 0.001$.
\footnote{Note that the lack of a detected AGN
in most high-redshift galaxies is not by itself an argument against
a duty cycle of unity for AGN with luminosities $-25<M_{1350}<-19$.
These AGN could shine exclusively within the most massive galaxies,
leaving the less massive galaxies with AGN that are undetectably faint.
}
Taking these arguments
into account would bring the high and low luminosity duty cycles
closer together in Figure~\ref{fig:r0dutycycle}.

Second, the physical interpretation is not straightforward.
Recall that we have defined the duty cycle for
the absolute magnitude range $M_{\rm lo}<M<M_{\rm hi}$ 
as the ratio of the number density of AGN with those
magnitudes to the number density of halos that can host them.
In the appendix we show that this duty cycle would be independent
of magnitude if blackholes accreted only at the Eddington rate,
were not obscured by dust, and had masses that followed a
tight power-law correlation with the total 
masses $M_h$ of galaxies that contain them.
The duty cycle would decrease at large luminosities
if brighter AGN were more heavily obscured, if black hole masses
fell below the predictions of the
$M_{\rm BH}$--$M_h$ correlation at very large $M_h$, or
if anything (e.g., complicated light curves) gave a broad range
of luminosities $L$ to the AGN that lie within halos of
a given mass $M_h$.  Each of these is expected theoretically
(e.g., Hopkins et al. 2005).  The apparent decrease of duty cycle
at large luminosities presumably results from a combination of
physical effects, and our observations do not identify which
is dominant among them.

Finally,
our result was derived from a small survey designed
for other purposes.  Most of the brightest AGN lay behind the
survey galaxies, not in their midst, reducing the number of
galaxy-AGN pairs and increasing the uncertainty in $r_0$.
A large, optimized survey
could easily
shrink the error bars several fold.
The only useful contribution of this paper may be its
demonstration that a definitive measurement is within easy reach.

\bigskip
\bigskip
KLA would like to thank
L. Ho, L. Hernquist, L. Ferrarese,
and J. Kollmeier for many interesting conversations
and an anonymous referee for encouraging us to discuss
the physical interpretation of the duty-cycle.
Our collaborators in the Lyman-break survey did most of the work
in taking and reducing these data.  We are grateful that
they let us proceed with the analysis.
This research has made use of the NASA/IPAC Extragalactic Database (NED) 
which is operated by the Jet Propulsion Laboratory, California 
Institute of Technology, under contract with the National Aeronautics 
and Space Administration.

\appendix

\section{PHYSICAL INTERPRETATION OF THE DUTY CYCLE}
We discuss three simple models for AGN evolution that may
help illustrate the physical meaning of the duty cycle.

Suppose first that black hole mass $M_{\rm BH}$ is tightly correlated with
total galaxy mass $M_h$ at all times, that
the correlation has the form $M_{\rm BH}\propto M_h^\alpha$, 
that AGN are unobscured by dust,
and that black holes 
radiate either at the Eddington rate $L_{\rm Edd}(M_{\rm BH})$
or not at all, gaining their mass in a few short accretion
episodes separated by long periods of quiescence.
The duty cycle would then be independent of AGN magnitude,
as can be seen with the following argument.

\begin{figure}
\plotone{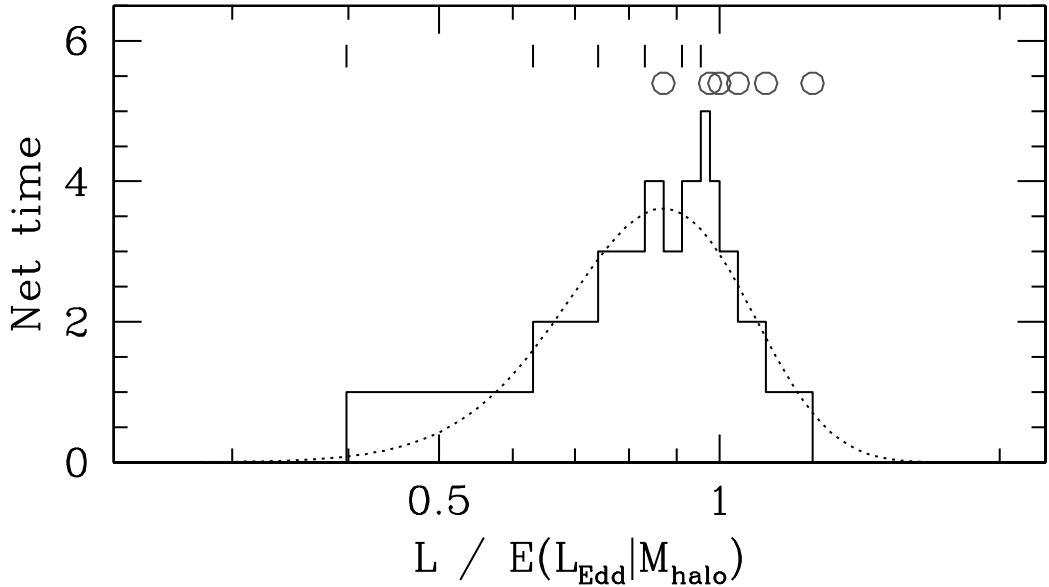}
\caption{
\label{fig:halokernel}
Net time spent at a given luminosity for
an ensemble of 6 blackholes in the first toy model
considered in the appendix.  We assume that these black holes radiate at the
Eddington luminosity and lie inside 6 halos of equal mass $M_h$.  
If the blackholes have
the initial luminosities that are marked with vertical lines,
and they grow until the luminosities have reached the final
values marked with circles, then the total amount of time the 6 AGN
spent radiating at a given luminosity is shown by the solid histogram.
The distribution for {\it all} AGN in halos of mass $M_h$,
not just these 6 AGN, might look more like the dotted curve
in the background.  If black hole and halo mass are tightly
correlated and all accretion is at the Eddington rate, this function
has to be narrow.
The units on the $x$ axis are normalized
to $E(L_{\rm Edd}|M_h)$, the mean Eddington luminosity of all AGN
in halos of mass $M_h$; units on the $y$ axis are arbitrary.
}
\end{figure}

Begin by considering the evolution
of a black hole inside a single dark matter halo of given mass $M_h$.
When the halo forms in the very early stages of a merger of
two smaller halos, its black hole mass\footnote{which may initially
be divided among two black holes; since the Eddington luminosity
of black hole of mass $2M$ is equal to the sum of
the Eddington luminosities two black holes each of mass $M$,
this does not affect our argument} 
may initially be smaller than the mean mass implied by
the $M_{\rm BH}$--$M_h$ correlation, but by the time the halo
is destroyed by mergers, roughly one Hubble time later 
(Martini \& Weinberg 2001), the black hole must have grown enough
to fall on the correlation.  Otherwise the correlation 
could not be satisfied by the ensemble of all halos.
Since accretion at the Eddington rate produces exponential growth,
the black hole will spend equal amounts of time in each octave
of luminosity as it grows from its initial mass $M_i$ to its final mass $M_f$;
if one were to plot the amount of time spent in each logarithmic
interval of luminosity $L$, it would be constant 
for $L_{\rm Edd}(M_i)<L<L_{\rm Edd}(M_f)$ and 0 elsewhere.
This is equally true if the growth occurs in many discrete episodes of accretion
or in a single burst.
Now consider a plot of total elapsed time versus luminosity for the black holes
within $N$ randomly chosen 
halos of the same mass $M_h$.  It would be the superposition of $N$ boxcars
with random left and right edges, producing an overall shape
that is peaked near the Eddington luminosity of the typical black hole
associated with halos of mass $M_h$.
Figure~\ref{fig:halokernel} shows an example for $N=6$.  
The same plot for the ensemble of
all halos of mass $M_h$ would be a smoother realization of a similar
function.  Call this plot the kernel.  
Since the number 
of AGN we observe with a given luminosity
is proportional to the net time AGN spend at that luminosity,
the kernel is the AGN luminosity distribution we would observe if
the universe consisted solely of halos with mass $M_h$.  
The width of the kernel depends on how far the
initial and final black hole masses stray from
the expectation value $E(M_{\rm BH}|M_h)$, but it must
be very narrow compared to the multi-decade width of the halo mass 
distribution.  Otherwise 
our assumption of a tight $M_{\rm BH}$--$M_h$ correlation would
be violated.  
The AGN within a narrow range of luminosity 
therefore must lie inside halos with a narrow range of masses.
Our definition of duty cycle for $L_{\rm min}<L<L_{\rm max}$
is the number density of AGN within that range of luminosity divided
by the number density of halos that can host them.
In this scenario, 
it is equal to the time required for the AGN's luminosity to grow from
$L_{\rm min}$ to $L_{\rm max}$ if it is accreting at the Eddington rate
divided by the halos' mean lifetime.  The numerator is independent
of halo mass for logarithmic luminosity intervals, and the
denominator depends extremely weakly on halo mass (Martini \& Weinberg 2001).  
Therefore the duty cycle in logarithmic luminosity bins
should be nearly independent of halo mass or AGN
luminosity.

\begin{figure}
\plotone{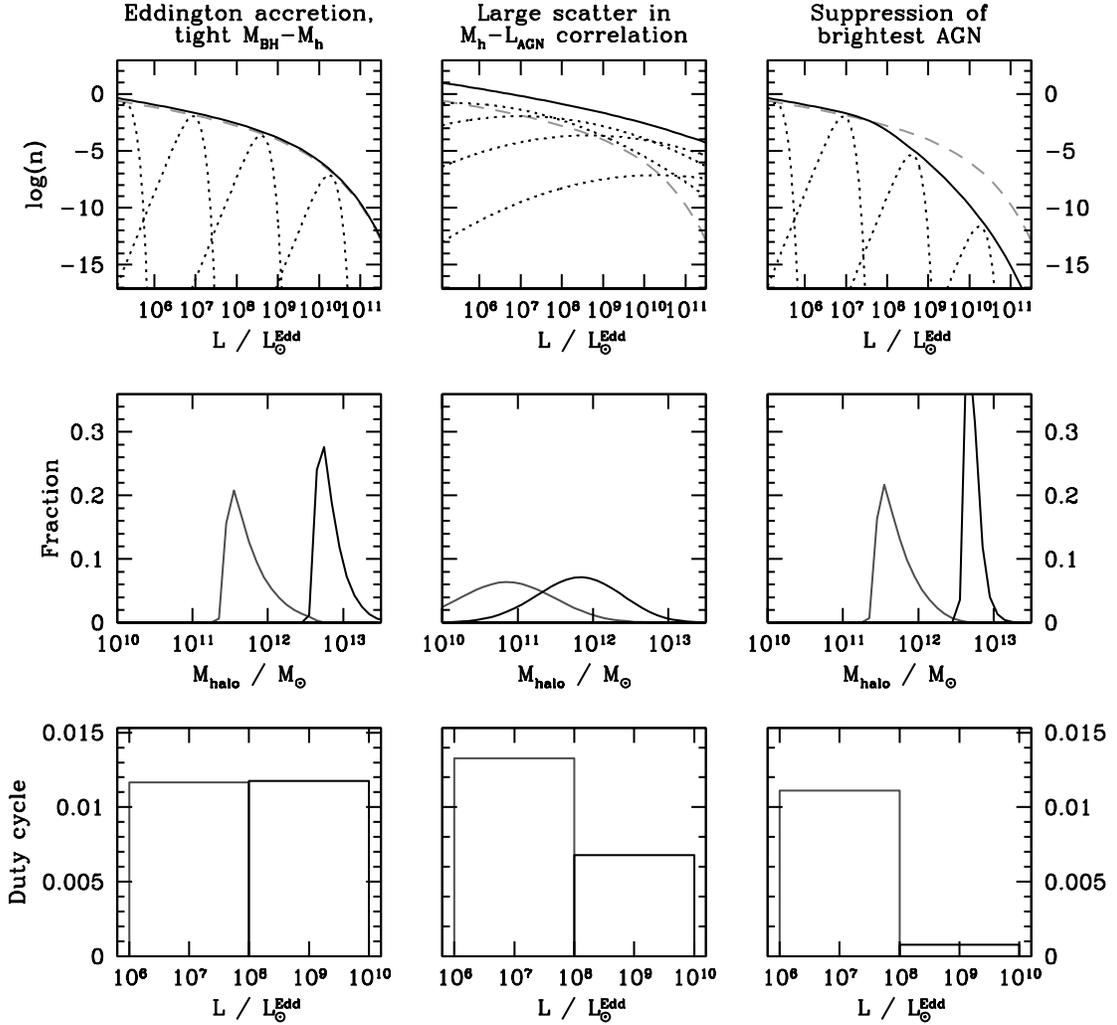}
\caption{
\label{fig:appendix}
Dependence of duty cycle on luminosity for three toy models.
We generated an ensemble of simulated central black holes 
from a Press-Schechter mass function by associating each
halo of mass $M_h$ with a blackhole of expected mass
$M_{\rm BH}/M_\odot = 10^7(M_h/10^{12}M_\odot)^{1.65}$
(Ferrarese 2002), then
associated each blackhole with a luminosity under different
assumptions for the three models.
Each column shows results for one model.
Left panels assume Eddington accretion and a tight correlation
of $M_h$ and $M_{\rm BH}$.  Middle panels
assume large scatter in the relationship between $M_h$ and 
AGN luminosity $L$.
Right panels assume Eddington accretion but stunt the
growth of blackholes in the most massive halos.
Units on the $y$ axis are arbitrary in all panels.
{\it Top panels:}
Dashed lines show the luminosity
distribution that would have resulted if each black hole 
had exactly the mean mass predicted by the $M_{\rm BH}$--$M_h$
relationship and radiated at
its Eddington luminosity.  
Instead we assumed that each halo's AGN luminosity
had some scatter around its expectation value.
Dotted lines indicate the assumed luminosity distribution
for halos of mass $10^{12}, 10^{13}, 10^{14}, 10^{15} M_\odot$.
They show the assumed scatter in the $M_h$--$L$ relationship,
which is large for the middle model and equally small
for the left and right models.  The solid lines
show the implied AGN luminosity function.
(More-realistic models would keep the AGN luminosity function
fixed to the observations by adjusting other parameters,
but these simple models are sufficient to illustrate our point.)
{\it Middle panels:}
Distribution of mass for halos that host AGN with luminosities
$10^6L^{\rm Edd}_\odot < L < 10^8L^{\rm Edd}_\odot$ 
and
$10^8L^{\rm Edd}_\odot<L<10^{10}L^{\rm Edd}_\odot$.
Here $L^{\rm Edd}_\odot$ is the Eddington luminosity of
a solar-mass blackhole.
{\it Bottom panels: }
Derived duty cycle for AGN in the same luminosity ranges.
The duty cycles were estimated by dividing the number
of AGN in the luminosity range by the number density of halos
more massive than the mean shown in the middle panels.
These panels show that
the duty cycle will decrease at high luminosities if there is
significant scatter in the $M_h$--$L$ relationship or
if black hole accretion is suppressed in the most
massive halos.  Increases in dust obscuration with luminosity
can also reduce the duty cycle at high $L$,
but we judged this effect too obvious to illustrate here.
}
\end{figure}

To check this claim, we generated an ensemble of simulated AGN
by starting with an ensemble of halos following a Press-Schechter
mass function ($\Omega_M=0.3$, $\Omega_\Lambda=0.7$, $\Gamma=0.2$,
$\sigma_8=0.9$, $z=2.5$), assigning each halo an expected central blackhole
mass with the relationship $M_{\rm BH} = 10^7 (M_h/10^{12} M_\odot)^{1.65}$
(Ferrarese 2002), and giving each black hole a luminosity
equal to the Eddington luminosity of the expected mass times a random
number drawn at random from the kernel (Figure~\ref{fig:halokernel}).
This resulted in the AGN luminosity distribution shown in the upper left
panel of Figure~\ref{fig:appendix}.  The distribution of halo masses for
AGN with luminosities $L_{\rm Edd}(10^6M_\odot)<L<L_{\rm Edd}(10^8M_\odot)$
and $L_{\rm Edd}(10^8M_\odot)<L<L_{\rm Edd}(10^{10}M_\odot)$
is shown in the middle left panel.  The bottom panel shows
the inferred duty cycle in these luminosity ranges, i.e., 
the ratio of AGN number density in each luminosity range to the
number density of halos more massive than the mean associated
halo mass shown in the middle left panel.  This is roughly the duty cycle
that would be estimated with the approach we adopted above.
It is the same for the
two logarithmic luminosity ranges, as expected.

The scenario can be altered in two ways to make the duty cycle decrease
at larger luminosities.  

The first is to increase the number density of
halos that can host the brightest AGN.  For a fixed halo mass distribution,
this can be accomplished by relaxing our assumption
that accretion occurs only at the Eddington rate or by increasing
the scatter in the $M_{\rm BH}$--$M_h$ relationship.
Either increases the scatter in the relationship
between $M_h$ and $L$, raising
the probability that a high luminosity AGN
resides within a low mass halo.
The middle column of Figure~\ref{fig:appendix} shows one example
of how a broad distribution of $L$ at fixed $M_h$ makes the
duty cycle depend on luminosity.  

The second way is to reduce the lifetimes of the brightest AGN.
If the $M_{\rm BH}$--$M_h$ correlation is a tight power-law and
all accretion is at the Eddington rate, then we can adjust
neither the mean accreted mass for blackholes in the most massive
halos nor the rate at which accretion occurs.  In this case
lifetimes of the brightest AGN can be reduced only by making
them heavily obscured while they accrete most of their mass.
The lifetimes can also be reduced, even for unobscured Eddington-rate
accretion, if we change the form of the $M_{\rm BH}$--$M_h$ relationship.
One change 
seems well motivated:  letting 
it break down for halos with super-galactic masses.
Ferrarese's (2002) relationship $M_{\rm BH}\sim 10^7 (M_h/10^{12} M_\odot)^{1.65} M_\odot$ predicts that local clusters of mass $10^{15}M_\odot$ should
contain $10^{12}M_\odot$ central blackholes, for example, but there
is no evidence that these ultra-massive blackhole exist. 
It seems more likely that blackhole formation becomes as suppressed
as star-formation in halos with mass $M_h\gg 10^{13}M_\odot$.
Suppressing or obscuring 
the brightest AGN can make the duty cycle depend
strongly on luminosity, as Figure~\ref{fig:appendix} shows.

If additional observations confirm the decrease in duty cycle
at high luminosities, some combination of these effects would presumably 
be responsible.

\end{document}